\documentclass[sigconf,10pt]{acmart}

\usepackage{array}
\usepackage{subcaption}
\usepackage[english]{babel}
\usepackage{enumitem}

\ifdefined\FivePage

\setcopyright{acmcopyright}
\copyrightyear{2019}
\acmYear{2019}
\setcopyright{acmcopyright}
\acmConference[SYSTOR '19]{The 12th ACM International Systems and Storage Conference}{June 3--5, 2019}{Haifa, Israel}
\acmBooktitle{The 12th ACM International Systems and Storage Conference (SYSTOR '19), June 3--5, 2019, Haifa, Israel}
\acmPrice{15.00}
\acmDOI{10.1145/3319647.3325837}
\acmISBN{978-1-4503-6749-3/19/06}

\else
\renewcommand\footnotetextcopyrightpermission[1]{} 
\setcopyright{none}

\fi

\newcommand{\ampm}{AM-PM}
\newcommand{\coll}{Collector}

\begin{document}

\date{}

\title{
Time-Multiplexed Parsing in \\ Marking-based Network Telemetry 
\ifdefined\FivePage
\else
\textsuperscript{\ensuremath\ast}
\fi
}

\author{
Alon Riesenberg\textsuperscript{\ensuremath\dagger}, Yonnie Kirzon\textsuperscript{\ensuremath\dagger}, Michael Bunin\textsuperscript{\ensuremath\dagger}, 
\\
Elad Galili\textsuperscript{\ensuremath\dagger}, Gidi Navon\textsuperscript{\ensuremath\bullet}, Tal Mizrahi\textsuperscript{\ensuremath\diamond}\textsuperscript{\ensuremath\dagger}
\\ 
\textsuperscript{\ensuremath\dagger}Technion --- Israel Institute of Technology, \textsuperscript{\ensuremath\bullet}Marvell Semiconductors 
\\
\textsuperscript{\ensuremath\diamond}Huawei Network.IO Innovation Lab
}        

\maketitle

\thispagestyle{empty}

\ifdefined\TwoPage
\textbf{Abstract.}
\else
\section*{abstract}
\fi
Network telemetry is a key capability for managing the health and efficiency of a large-scale network. Alternate Marking Performance Measurement (AM-PM) is a recently introduced approach that accurately measures the packet loss and delay in a network using a small overhead of one or two bits per data packet. 
This paper introduces a novel time-multiplexed parsing approach that enables a practical and accurate implementation of AM-PM in network devices, while requiring just a single bit per packet. Experimental results are presented, based on a hardware implementation, and a software P4-based implementation. 

\ifdefined\FivePage
\else
{\let\thefootnote\relax\footnotetext{
\textsuperscript{\ensuremath\ast}Accepted to ACM SYSTOR, 2019. This manuscript is a pre-published extended version.
}}
\fi

\section{Introduction}
\ifdefined\TwoPage
\else
In the past decade the scale and performance of high-speed networks has increased dramatically, while network operators have been constantly striving for flexible, easy-to-manage and automated networks. One of the common threads that run through these different requirements is network telemetry; the ability to accurately measure the performance of the network in real-time.

Network telemetry standards and protocols have been in deployment for over two decades. In the last few years new approaches have been introduced, which piggyback telemetry information onto data packets, allowing fine-grained per-hop and per-packet measurement. 
In-situ OAM (IOAM)~\cite{IOAM} and In-band Network Telemetry (INT)~\cite{kimband,INT} are two in-band telemetry approaches that have been gaining momentum.

Another telemetry method that is being pursued by the Internet Engineering Task Force (IETF) is the Alternate Marking Performance Measurement (AM-PM) method~\cite{AltMark}, which is a lightweight in-band measurement approach that requires one or two bits per data packet. 
\fi

\ampm\ ~\cite{AltMark} is a performance measurement method that provides accurate and reliable in-band measurement with a negligible overhead of one or two bits in the header of every data packet. \ampm\ is currently under development in the IETF; it was introduced in a recently published RFC~\cite{AltMark}, and is being considered in the context of various encapsulations, including Geneve, SFC NSH, BIER, MPLS, and QUIC. Notably, \ampm\ can also be deployed over IPv4 or IPv6 by using reserved values of fields in the IP header. At least one known deployment~\cite{AMPMCom}, in Telecom Italia's production network, uses reserved values in the IP header for \ampm.

\ifdefined\TwoPage
\begin{figure}[!t]
	\centering
  \fbox{\includegraphics[width=.46\textwidth]{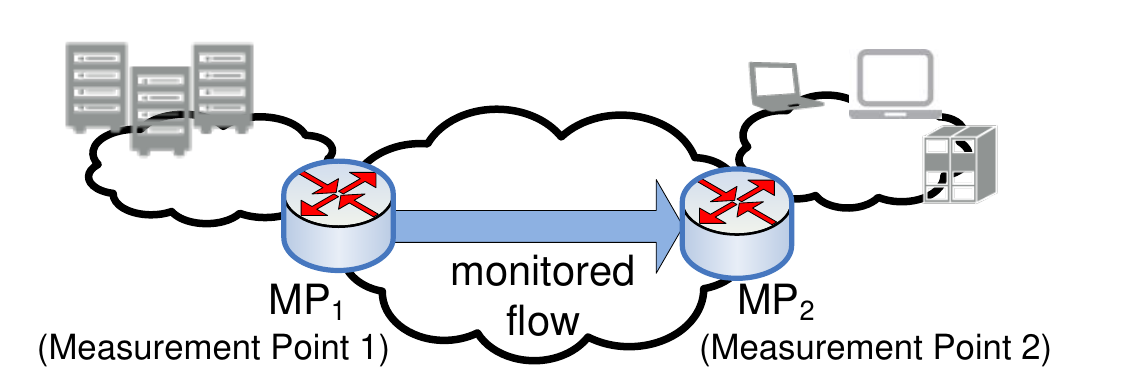}}
	\captionsetup{justification=raggedright}
  \caption{Two Measurement Points, $MP_1$ and $MP_2$.}
  \label{fig:Measurement}
  \vspace{-5mm}
\end{figure}
\else
\begin{figure}[!b]
	\centering
  \fbox{\includegraphics[width=.46\textwidth]{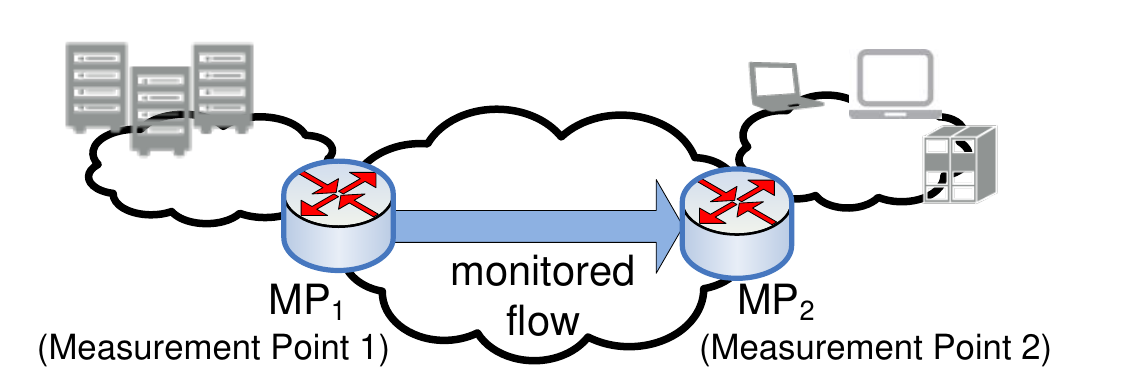}}
	\captionsetup{justification=raggedright}
  \caption{Performance measurement between two Measurement Points, $MP_1$ and $MP_2$.}
  \label{fig:Measurement}
\end{figure}
\fi

\ifdefined\TwoPage
In this paper we introduce a novel time-multiplexed parsing approach that enables the implementation of 
\ampm\ in network devices such as switches, routers, and NICs. 
Notably, this paper confirms that \ampm\ can perform accurate measurement with a single marking bit per packet without significant implementation complexity.

Experimental results are presented, based on a hardware implementation using a Marvell Prestera switch silicon, and a software implementation in the P4~\cite{bosshart2014p4} programming language, which is publicly available as open source~\cite{code}.
\else
The main contributions of this work are as follows:

\begin{itemize}[leftmargin=*]
  \item We introduce a novel time-multiplexed parsing approach that enables the implementation of \ampm\ in network devices such as switches, routers, and NICs.
	\item The abstractions required to implement marking-based telemetry are analyzed from a silicon design perspective, as well as from a data plane programming perspective. Two primitive abstractions that are key for implementing marking-based telemetry are presented, and their application to \ampm\ is discussed and demonstrated.
	\item Experimental results are presented, based on a hardware implementation using a Marvell Prestera switch silicon, and a software implementation in the P4~\cite{bosshart2014p4} programming language.
	\item Our P4-based implementation is publicly available as open source~\cite{code}.
\end{itemize}
\fi

\ifdefined\LessText
Due to space limits, some of the detailed analysis is presented in an extended technical report~\cite{MarkingBasedTech}.
\fi

\section{\ampm\ in a Nutshell}
\ifdefined\TwoPage
\else
This section provides a short overview of \ampm\ ~\cite{AltMark,AMPMCom}.
\fi

As illustrated in Fig.~\ref{fig:Measurement}, \ampm\ is used between two or more Measurement Points (MP) in the network. 
\ifdefined\TwoPage
\else
MPs can be hosts, servers, or network devices. 
\fi
The data traffic between the MPs carries one or two \emph{marking bits}. 
This implies that the initiating MP ($MP_1$ in the figure) assigns the value of the marking bit(s), and the terminating MP ($MP_2$) acts upon their value. If $MP_2$ is not the final destination of the packet it clears the value of the marking bits to zero, so as not to affect successive nodes that do not take part in the measurement. 
\ifdefined\TwoPage
\else

\fi
The marking bits are used for signaling and coordinating measurement events between the MPs. A measurement event is detected when either a `step' or a `pulse' in the value of the marking bit is detected.
\ifdefined\TwoPage
\else
If a tunnel encapsulation is used between the MPs, the marking bits are part of the encapsulation header, and thus the marking bits are removed when the terminating MP strips the encapsulation header from the packet.

A measurement may be performed among multiple MPs, as further discussed in~\cite{ietf-ippm-multipoint-alt-mark}. For ease of presentation this paper focuses on a point-to-point measurement between two MPs.
\fi

\begin{figure}[!b]
\ifdefined\TwoPage
\vspace{-5mm}
\fi
	\centering
  \fbox{\includegraphics[width=.46\textwidth]{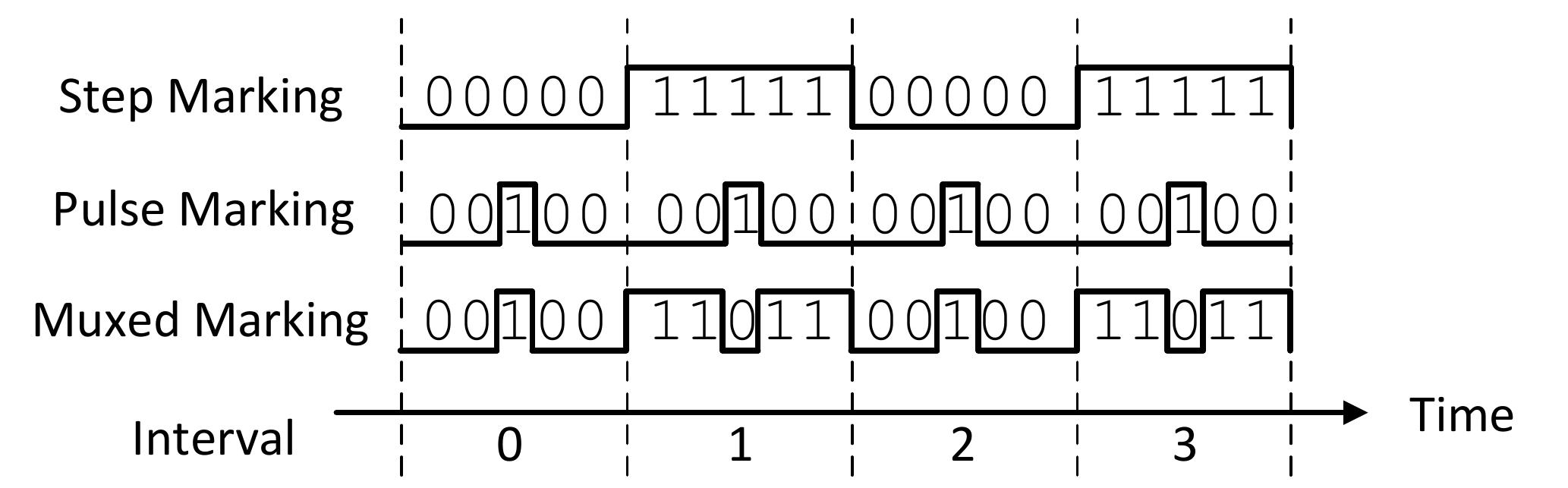}}
	\captionsetup{justification=raggedright}
  \caption{\ampm\ marking bits.}
  \label{fig:MarkingBits}

\end{figure}

A marking bit can be used in one of three possible ways, as illustrated in Fig.~\ref{fig:MarkingBits}:
\ifdefined\TwoPage
(i) the marking bit is toggled periodically, and a `step' indicates an event, (ii) the marking bit is assigned a `pulse' value once pet time interval, or (iii) the value of the marking bit is toggled periodically, and there is an additional pulse in during each interval. The muxed marking bit can be viewed as the result of an exclusive or (XOR) between a pulse bit and a step bit, thus multiplexing two marking bits onto a single bit.
\else

\subsubsection*{Step}
The marking bit is assigned a periodically alternating value. Thus, the marking bit divides the traffic into consecutive blocks of data. A `step' is detected when the value of the marking bit is toggled.

\subsubsection*{Pulse}
The marking bit is assigned a `pulse' value once per time interval. The pulse value indicates to the measurement points that the marked packet is a reference for the measurement; for example, the `pulse' packet indicates that all devices should capture the timestamp of this packet.

\subsubsection*{Multiplexed}
This approach~\cite{Compact} combines the step and pulse approaches; the value of the marking bit is toggled periodically, and there is an additional pulse in the middle of each interval. The muxed marking bit can be viewed as the result of an exclusive or (XOR) between a pulse bit and a step bit, thus multiplexing two marking bits onto a single bit.

\fi

\ifdefined\TwoPage
The marking bit(s) is used to signal measurement events between the two measurement points, allowing a coordinated and consistent measurement. 
The main advantage of \ampm\ is that it provides accurate loss and delay measurement at the cost of one or two bits per data packet.
\else
The simplest variant of \ampm\ is called the \emph{double marking} method, in which two marking bits are used: a `step' marking bit is used for Loss Measurement (LM), and a `pulse' bit is used for Delay Measurement (DM). The \emph{step bit} marks the data packets with one of two `colors', 0 or 1. Each of the MPs maintains two counters,\footnote{Two counters per flow.} one per color.  At the end of each interval the counter values can be collected by a central \coll\ \footnote{A collector may be an analytics server, a Network Management System (NMS) or a Software Defined Network (SDN) controller.} and analyzed. For example, at the end of a `0' interval, counter 0 is collected from both MPs, and the packet loss during the interval can be computed by comparing the counter values of the two MPs. Notably, the use of two colors intermittently implies that each counter is exported when it is not currently in use, and thus its value is stable, allowing a consistent snapshot of the counter across the MPs. The \emph{pulse bit} is used to indicate specific packets that are used for DM; both MPs measure the timestamp of that packet, and export it to the \coll, which can compute the path delay by comparing the two timestamps.

The main advantage of \ampm\ is that it provides accurate loss and delay measurement at the cost of one or two bits per data packet.
\fi

\ifdefined\TwoPage
\section{Time-Muxed Parsing}
\else
\section{Marking-based Telemetry: Design and Implementation}
\fi
\ifdefined\TwoPage
We introduce a novel time-muxed parsing approach, in which we divide time into slots, and the semantic interpretation of the marking bit is a function of the time slot.
Our analysis focuses on the muxed marking method, which muxes the step and pulse bits onto a single bit. Note that it is possible to implement the two simpler approaches using a subset of the principles described here.  
The advantage of muxed marking is that it requires just a single bit per packet on the wire.  

We use two basic abstractions in our implementation. These two basic abstractions are supported in off-the-shelf network device silicons, as well as in the P4 programming language: (i) \emph{Time-as-a-match}: If the time of every packet is measured as the packet is received, then this timestamp can be used as a match criterion in the device's match-action lookup. Specifically, the timestamp should be used in a ternary lookup, in which most of the timestamp bits are masked, defining a match rule that is matched periodically. This is a special case of the approach suggested in~\cite{Infocom-TimeFlip}. (ii) \emph{Stateful behavior}: The packet processing decision can be based not only on the packet header, but also on a state. For example, stateful behavior allows distinction between the first packet of an interval, and other packets, thus enabling the `pulse' abstraction. 
\else

This section introduces a design and implementation of \ampm. One of the main challenges in the context of implementing marking-based telemetry is the non-trivial functionality that is required in this telemetry method. As illustrated in Fig.~\ref{fig:Measurement}, $MP_1$ is the node that initiates the measurement. Thus, it assigns the marking bit(s). In order to implement step marking, $MP_1$ is required to periodically toggle the marking bit. Pulse marking requires $MP_1$ to select one of the packets in each interval and process it differently than other packets. 

\subsection{Primitive Abstractions}
We now discuss two primitive abstractions that can be used to implement \ampm. Fortunately, these two basic abstractions are supported in off-the-shelf network device silicons, as well as in the P4 programming language, thus enabling the implementations described in Section~\ref{EvalSec}.

\subsubsection*{Time bit as a match criterion.} 
If the time of every packet is measured as the packet is received, then this timestamp can be used as a match criterion in the device's match-action lookup. Specifically, the timestamp should be used in a ternary lookup, in which all but one of the timestamp bits are masked, defining a match rule that is matched periodically. This is a special case of the approach suggested in~\cite{Infocom-TimeFlip}. In hardware-based network devices, this requires the ability to include the packet's timestamp in a ternary match, such as a Ternary Content Addressable Memory (TCAM) lookup. This programming abstraction is provided by P4, as the packet's ingress time is included in the P4 intrinsic metadata, allowing the timestamp to be used in ternary matches~\cite{P4Spec14,PSA}.
	
\subsubsection*{Stateful behavior.} 
The packet processing decision can be based not only on the packet header, but also on a state. For example, stateful behavior allows distinction between the first packet of an interval, and other packets, thus enabling pulse marking. Stateful behavior can be implemented in hardware devices by maintaining a per-flow state, and applying an action that is based on this state. P4 supports stateful memories that consist of registers that can be used in the match-action procedure.
\fi

\ifdefined\TwoPage
\else
\subsection{Step Marking Implementation}
Step marking requires periodic behavior, such that the value of the marking bit is toggled in every time interval. The time-bit-as-a-match abstraction can be used to implement a periodic time range. For example, assume that the timestamp consists of two fields, a \emph{Seconds} field and a \emph{Second Fraction} field. The least significant bit of the Seconds field is by definition toggled every second. Thus, a ternary match rule that considers the least significant bit of the Seconds field and masks the rest of the bits (see Eq.~\ref{eq:TimeFlip}) defines a periodic match with an interval of 1 second. Masked bits are denoted by `$\ast$'.

\begin{equation}
\label{eq:TimeFlip}
\end{equation}

\vspace{-11mm}

\begin{equation}
\label{eq:DummyTimeFlip}
\nonumber
\includegraphics[width=.18\textwidth]{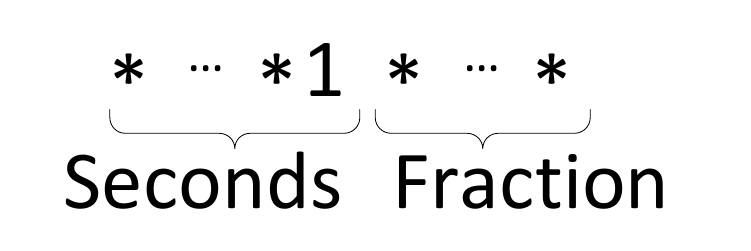}
\end{equation}

For each packet we denote the value of the time bit corresponding to the current packet by $TimeBit$, and $MarkBit$ indicates the value of the marking bit that is assigned to the packet by the initiating MP. These notations will be used throughout the section.

Continuing the example of Eq.~\ref{eq:TimeFlip}, the following match-action rules (Table~\ref{table:StepMatchAction}) can be used by the initiating MP to implement step marking for loss measurement. For example, the first rule implies that if $TimeBit$ is zero, then the marking bit is set to zero, and $counter0$ is used for counting.

\begin{table}[htbp]
    \begin{tabular}{| p{3.7cm}<{\centering} | p{3.7cm}<{\centering} |}
    \hline
    Match & Action  \\ \hline \hline
    $TimeBit = 0$ & $MarkBit = 0$, $counter0$  \\ \hline 
    $TimeBit = 1$ & $MarkBit = 1$, $counter1$  \\ \hline 
    \end{tabular}
    \caption{Step marking loss measurement example: the match-action table of the \emph{initiating} MP.}
    \label{table:StepMatchAction}
\end{table}

In this case, the match-action lookup at the terminating MP can be implemented by the rules of Table~\ref{table:StepTermMatchAction}.

\begin{table}[htbp]
    \begin{tabular}{| p{3.7cm}<{\centering} | p{3.7cm}<{\centering} |}
    \hline
    Match & Action  \\ \hline \hline
    $MarkBit = 0$ & $counter0$  \\ \hline 
    $MarkBit = 1$ & $counter1$  \\ \hline 
    \end{tabular}
    \caption{Step marking loss measurement example: the match-action table of the \emph{terminating} MP.}
    \label{table:StepTermMatchAction}
\end{table}

Generally speaking, time-bit-as-a-match can be used to implement periodic behavior, where the location of the unmasked bit determines the interval length.

\subsection{Pulse Marking Implementation}
In the pulse marking approach the initiating MP marks a single packet per interval as `1'. 
\ifdefined\FivePage
\else
This requires the initiator to be interval-aware, and in each interval we would like to implement the state machine of Fig.~\ref{fig:States}. `First' is the state in which the current packet is the first in the current interval, whereas `Not First' indicates the current packet is not the first in the interval. Upon the arrival of a packet, the state can be determined by comparing the current value of $TimeBit$ to the value of the time bit in the previous packet, denoted by $PrevTimeBit$.

\begin{figure}[htbp]

	\centering
  \fbox{\includegraphics[width=.3\textwidth]{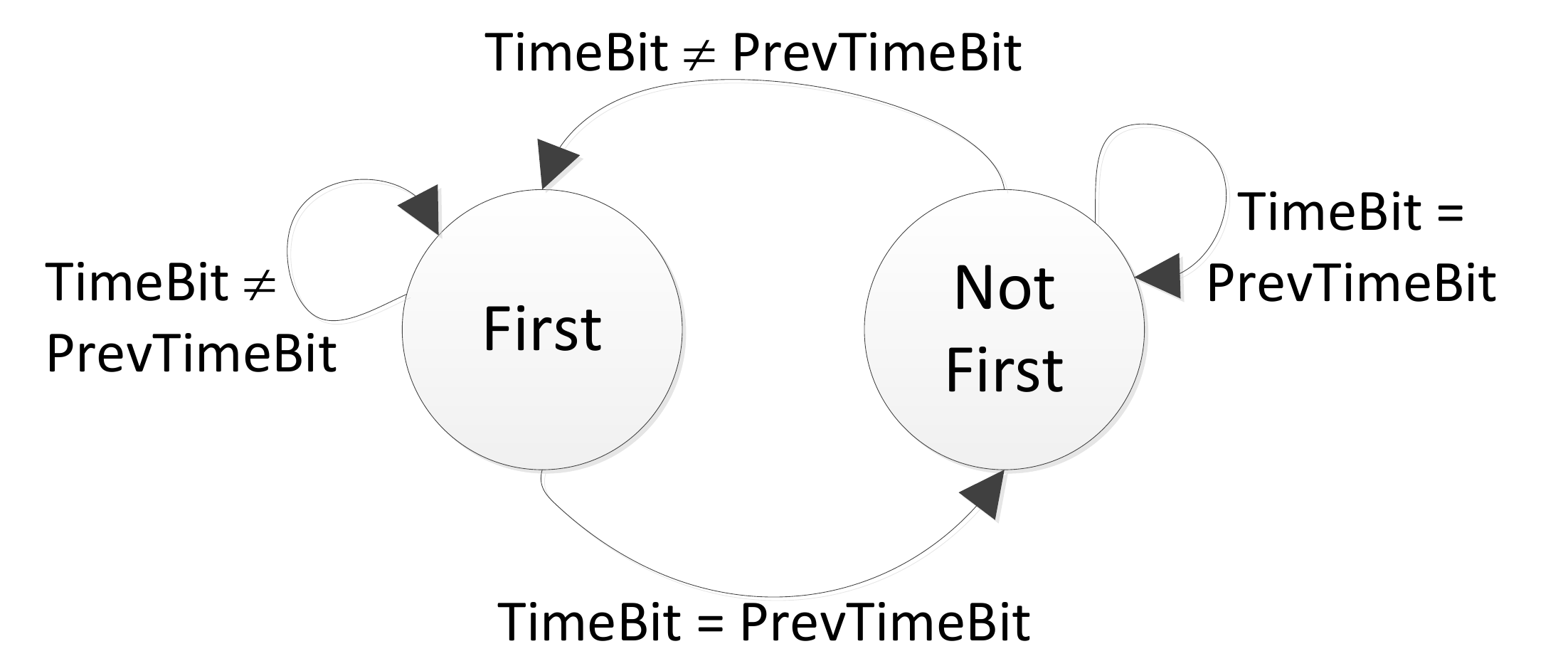}}
	\captionsetup{justification=raggedright}
  \caption{State diagram for detecting the first packet of the interval.}
  \label{fig:States}

\end{figure}
\fi

Pulse marking can be implemented using the time bit and a register, $Reg$, which represents the previous value of $TimeBit$. 
\ifdefined\FivePage
Further details are presented in the extended version of this paper~\cite{MarkingBasedTech}.
\else
As shown in Table~\ref{table:PulseMatchAction}, when the marking bit is set to `1' the initiating MP also records the timestamp of the packet, to be exported to the \coll.

\begin{table}[htbp]
    \begin{tabular}{| p{3.7cm}<{\centering} | p{3.7cm}<{\centering} |}
    \hline
    Match & Action  \\ \hline \hline
    $TimeBit = 1$, $Reg=0$ & $MarkBit=1$, $Reg=1$, $timestamp$ \\ \hline 
    $TimeBit = 1$, $Reg=1$ & $MarkBit=0$, $Reg=1$ \\ \hline 
    $TimeBit = 0$, $Reg=1$ & $MarkBit=1$, $Reg=0$, $timestamp$ \\ \hline 
    $TimeBit = 0$, $Reg=0$ & $MarkBit=0$, $Reg=0$ \\ \hline 
    \end{tabular}
    \caption{Pulse marking for delay measurement example: match-action table of the \emph{initiating} MP.}
    \label{table:PulseMatchAction}
\end{table}

The match rule at the terminating MP is simple (Table~\ref{table:PulseTermMatchAction}), as it only needs to monitor the value of the marking bit, and record the timestamp when its value is `1'.

\begin{table}[htbp]
    \begin{tabular}{| p{3.7cm}<{\centering} | p{3.7cm}<{\centering} |}
    \hline
    Match & Action  \\ \hline \hline
    $MarkBit=1$ & $timestamp$ \\ \hline 
    \end{tabular}
    \caption{Pulse marking for delay measurement example: match-action table of the \emph{terminating} MP.}
    \label{table:PulseTermMatchAction}
\end{table}

\fi

\subsection{Time-Multiplexed Parsing}
\label{MuxedSection}
\fi

We now introduce a novel time-muxed parsing approach, in which we divide time into slots, and the semantic interpretation of the marking bit is a function of the time slot.
Our analysis of the time-muxed approach focuses on the muxed marking method, which muxes the step and pulse bits onto a single bit. 
The advantage of muxed marking is that it requires just a single bit per packet on the wire, while providing the same measurement resolution and accuracy as the double marking approach.  


\ifdefined\TwoPage
Multiplexed marking combines the principles of step and pulse marking. We consider an example in which \ampm\ is run with an interval of 16~seconds. Thus, the fifth bit of the Seconds field, denoted by $Seconds[4]$, which is toggled every 16~seconds, and the periodic behavior of the `step' is implemented by a match rule that uses $Seconds[4]$, as shown in Eq.~\ref{eq:TimeFlipMuxed}.  
\else
Multiplexed marking combines the principles of step and pulse marking. We consider a typical case, in which muxed marking is used for both loss and delay measurement; step marking is used to determine which of the two counters is used ($counter0$ or $counter1$), and the pulse is used for indicating a timestamped packet. We consider an example in which \ampm\ is run with an interval of 16~seconds. 
As shown in Eq.~\ref{eq:TimeFlipMuxed}, the time bit is the fifth bit of the Seconds field, denoted by $Seconds[4]$, which is toggled every 16~seconds. The other bits are masked.
\fi

\begin{equation}
\label{eq:TimeFlipMuxed}
\end{equation}

\vspace{-11mm}

\begin{equation}
\label{eq:DummyTimeFlipMuxed}
\nonumber
\includegraphics[width=.18\textwidth]{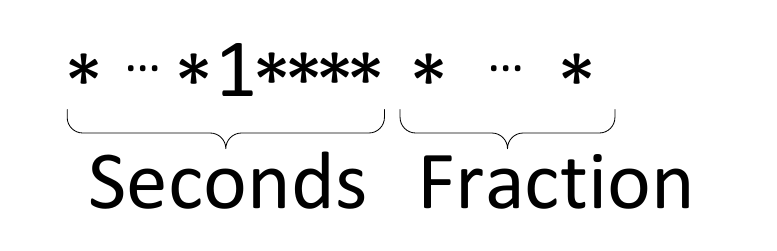}
\end{equation}

\begin{figure}[htbp]

	\centering
  \fbox{\includegraphics[width=.46\textwidth]{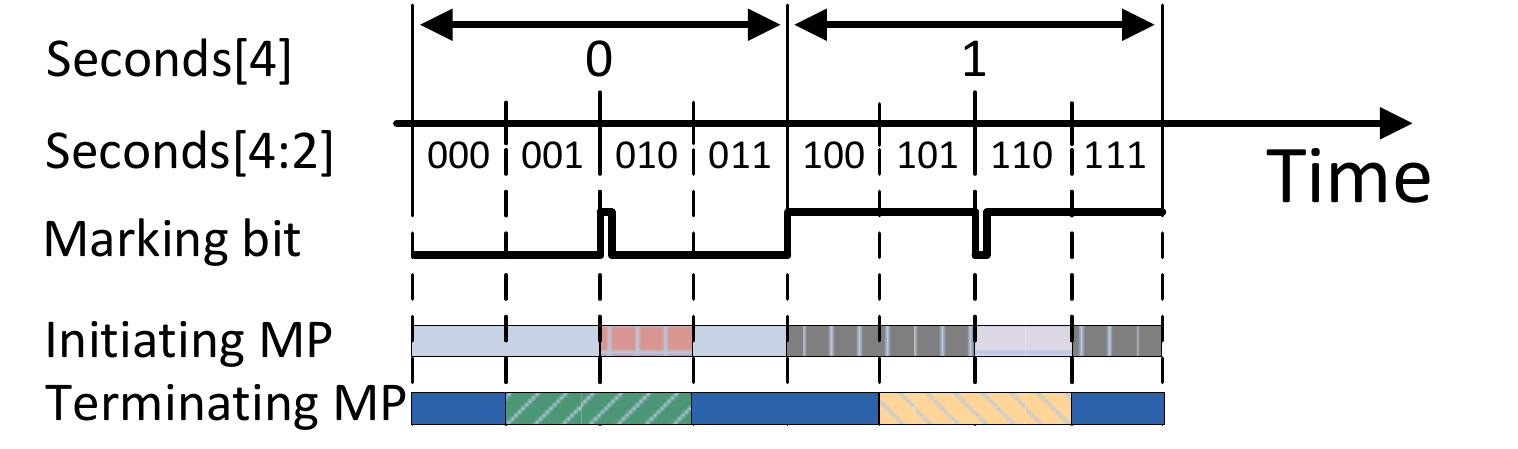}}
	\captionsetup{justification=raggedright}
  \caption{Time-multiplexed parsing: the marking bit has a different interpretation in each time slot.}
  \label{fig:TimeSlots}

\end{figure}

\ifdefined\TwoPage
In our time-muxed parsing approach we observe three bits from the timestamp field, $Seconds[4:2]$, dividing each 16-second interval into four sub-intervals, as shown in Fig.~\ref{fig:TimeSlots}.
\else
In order to implement muxed marking, we introduce a novel time-multiplexed matching approach. We observe three bits from the timestamp field, $Seconds[4:2]$, dividing each 16-second interval into four sub-intervals (each sub-interval is illustrated in a different color in Fig.~\ref{fig:TimeSlots}). 
\fi

In a nutshell, the `pulse' bit is invoked by $MP_1$ roughly in the middle of the interval, in order to minimize the potential confusion between a step and a pulse. Thus, $MP_1$ observes the third quarter of each interval, and triggers a pulse in the first packet of this sub-interval. At $MP_2$, the second and third quarter of each interval are used for detecting a pulse, whereas the rest of the time is used for detecting steps.

\ifdefined\TwoPage
\else

\ifdefined\FivePage
\else

Based on the intervals and sub-intervals defined by the three timestamp bits, the match-action rules at the initiating MP are specified in Table~\ref{table:MuxedSendMatchAction}. 

\begin{table}[htbp]
    \begin{tabular}{| p{3.7cm}<{\centering} | p{3.7cm}<{\centering} |}
    \hline
    Match & Action  \\ \hline \hline
    $TimeBits = 010$, $Reg=0$ & $MarkBit=1$, $Reg=1$, $counter0$, $timestamp$ \\ \hline 
    $TimeBits = 010$, $Reg=1$ & $MarkBit=0$, $Reg=1$, $counter0$ \\ \hline 
    $TimeBits = 0 {\ast} {\ast}$, $Reg={\ast}$ & $MarkBit=0$, $Reg=0$, $counter0$ \\ \hline 
    $TimeBits = 110$, $Reg=1$ & $MarkBit=0$, $Reg=0$, $counter1$, $timestamp$ \\ \hline 
    $TimeBits = 110$, $Reg=0$ & $MarkBit=1$, $Reg=0$, $counter1$ \\ \hline 
    $TimeBits = 1 {\ast} {\ast}$, $Reg={\ast}$ & $MarkBit=1$, $Reg=1$, $counter1$ \\ \hline 
    \end{tabular}
    \caption{Muxed marking for loss and delay measurement: match-action rules at the \emph{initiating} MP.}
    \label{table:MuxedSendMatchAction}
\end{table}

The corresponding rules at the terminating MP are specified in Table~\ref{table:MuxedRecMatchAction}.

\begin{table}[htbp]
    \begin{tabular}{| p{3.7cm}<{\centering} | p{3.7cm}<{\centering} |}
    \hline
    Match & Action  \\ \hline \hline
    $TimeBits = 001$, $MarkBit=1$ & $counter0$, $timestamp$ \\ \hline 
    $TimeBits = 010$, $MarkBit=1$ & $counter0$, $timestamp$ \\ \hline 
    $TimeBits = 101$, $MarkBit=0$ & $counter1$, $timestamp$ \\ \hline 
    $TimeBits = 110$, $MarkBit=0$ & $counter1$, $timestamp$ \\ \hline 
    $TimeBits = {\ast} {\ast} {\ast}$, $MarkBit=0$ & $counter0$ \\ \hline 
    $TimeBits = {\ast} {\ast} {\ast}$, $MarkBit=1$ & $counter1$ \\ \hline 
    \end{tabular}
    \caption{Muxed marking example for loss and delay measurement: match-action rules at the \emph{terminating} MP.}
    \label{table:MuxedRecMatchAction}
\end{table}

\fi
\fi

Time-multiplexed matching implicitly assumes that the initiating MP and the terminating MP are phase-synchronized at the slot level. It should be noted that muxed marking could be implemented differently without this assumption, by detecting a `step' in the marking bit value without considering the current time.

\ifdefined\FivePage
Further details about multiplexed marking, including a detailed description of the match rules, are presented in the extended version of this paper~\cite{MarkingBasedTech}.
\fi

\ifdefined\TwoPage
\else

\ifdefined\FivePage
\else
\subsection{Scalability}
Since \ampm\ will typically be applied on a per-flow basis, it is important for the implementation to be scalable. At a first glance, the previous subsection implies that \ampm\ may require $6$ match rule per flow. However, the efficiency can be significantly improved if two match-action lookups are used; the first lookup can be used to detect the timeslot and assign the marking bit value, while the second lookup can be used to classify the flow. Thus, the first lookup will only require a few global rules, while the second lookup will require two rules per flow, assuming that two counters per flow are used.\footnote{Note that if pulse marking is used for both loss and delay measurement, then only one counter per flow is required, and thus only one match rule per flow.}

\subsection{Performance Overhead}
The design and implementation of \ampm\ as described above can be applied to data traffic in full-wire-speed, without compromising the traffic bandwidth. However, since timestamps and counters are exported to a \coll, \ampm\ requires management overhead that is a function of the number of monitored flows and the measurement interval. The measurement interval that was evaluated in this work (Section~\ref{EvalSec}) was 1~second, but when highly granular measurement is desired a shorter interval may be used, yielding higher management overhead. It should be noted that the latter observation would similarly apply to any network telemetry approach, and is not specific to \ampm.

\subsection{Time synchronization}
\label{SyncSec}
\ampm\ can be implemented even if the clocks of the MPs are not synchronized. This is an important observation, as network devices do not necessarily have synchronized clocks. The P4 timestamp~\cite{P4Spec14,PSA} was intentionally defined in way that does not require the timestamp to be based on a synchronized time-of-day. For example, the timestamp in the P4 reference switch represents the time elapsed since the switch was powered up. The clocks of the MPs only need to be synchronized at the measurement interval level, which is a relaxed synchronization requirement; for example, an interval of 1~second requires the MPs to be synchronized with an accuracy on the order of 0.5~sec.

It should be noted that accurate time synchronization using the Precision Time Protocol (PTP)~\cite{IEEE1588} is a mature and commonly used technology that may often be available in networks that require performance measurement. We argue that there is a significant advantage if the time-bit-as-a-match relies on a globally synchronized time-of-day. One obvious advantage of synched clocks is that the one-way-delay between the two measurement points can be computed by simply comparing two timestamps measured in the same interval.\footnote{If the two MPs are not synchronized, it is possible to measure the two-way delay between them, assuming that the monitored traffic is bidirectional.} Another advantage for having (roughly) synchronized clocks is that the \coll\ can schedule periodic sampling of the devices' counters. For example, the \coll\ can have a periodic process that collects the counters of the previous interval in the middle of the current interval, thereby guaranteeing that the counters of the previous interval are currently not in use. Finally, the muxed marking approach that was presented in the previous section assumes that $MP_1$ and $MP_2$ are synchronized at the interval level.
\fi
\fi

\section{Evaluation}
\label{EvalSec}
We evaluated two implementations; a hardware implementation based on a Marvell switch silicon, and a P4-based software implementation~\cite{code}.
\ifdefined\LessText
We present a glimpse at some of the experimental results, showing the high accuracy of the delay measurement.
Detailed results are presented in~\cite{MarkingBasedTech}.

\begin{figure}[htbp]
	\centering

  \begin{subfigure}[t]{.24\textwidth}
	\centering
  \fbox{\includegraphics[height=.75\textwidth]{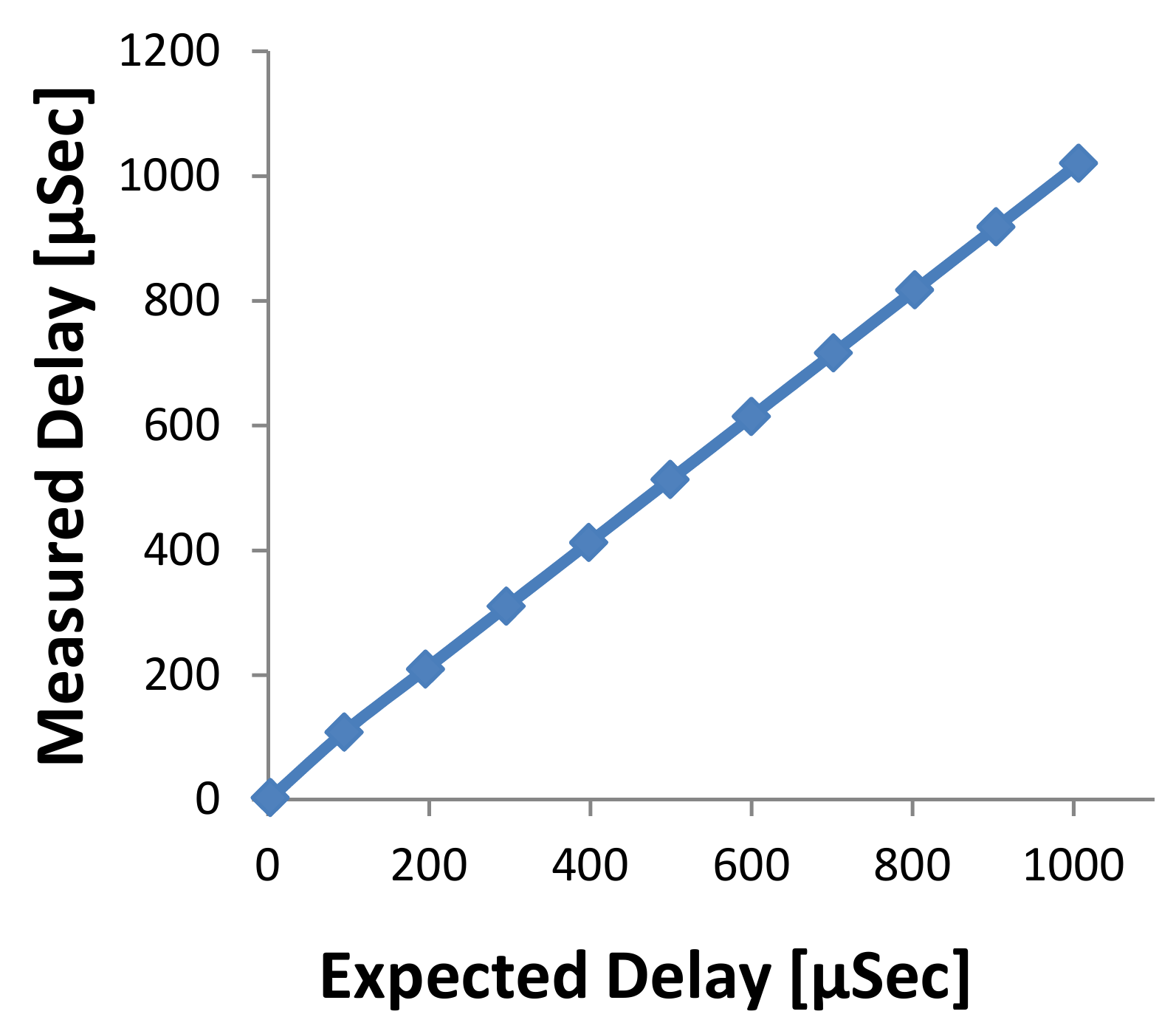}}
	\captionsetup{justification=centering}
  \caption{Delay in the HW implementation: \\ measured vs. expected.}
  \label{fig:ShortDemoDelay}
  \end{subfigure}%
  \begin{subfigure}[t]{.24\textwidth}
	\centering
  \fbox{\includegraphics[height=.75\textwidth]{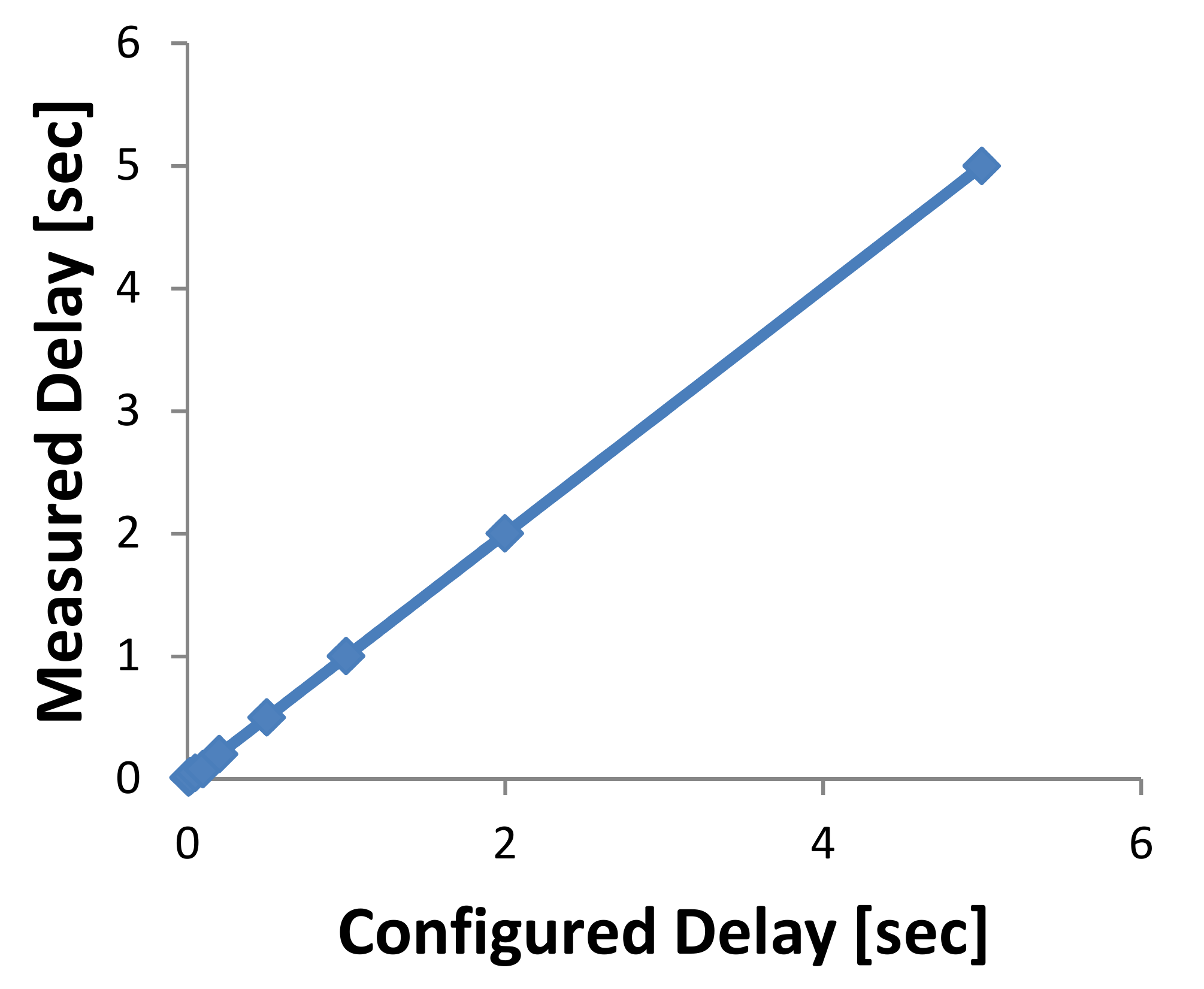}}
	\captionsetup{justification=centering}
  \caption{Delay in the SW implementation: \\ measured vs. expected.}
  \label{fig:ShortDelayResults}
  \end{subfigure}%

	\captionsetup{justification=raggedright}
  \caption{Experimental results of delay measurement.}
  \label{fig:DelayLoss}
\end{figure}

\ifdefined\TwoPage
\else

Due to space limits, some of the experimental results are presented in~\cite{MarkingBasedTech}.
\fi
\fi

\ifdefined\TwoPage
\else

\begin{figure}[htbp]
	\centering

  \begin{subfigure}[t]{.24\textwidth}
	\centering
  \fbox{\includegraphics[height=.85\textwidth]{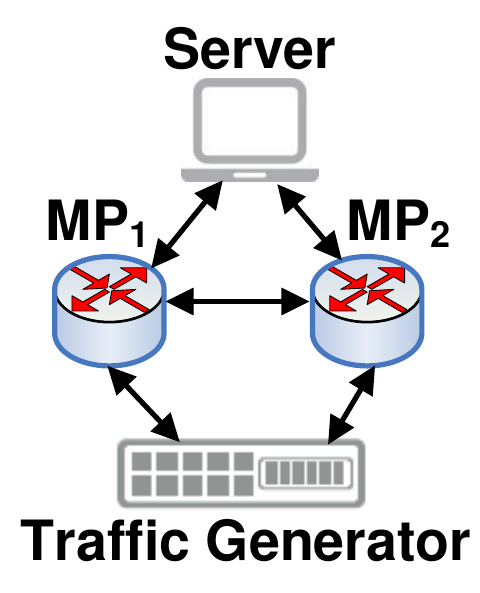}}
	\captionsetup{justification=centering}
  \caption{HW experiment setup.}
  \label{fig:ExperimentNetwork}
  \end{subfigure}%
  \begin{subfigure}[t]{.24\textwidth}
	\centering
  \fbox{\includegraphics[height=.85\textwidth]{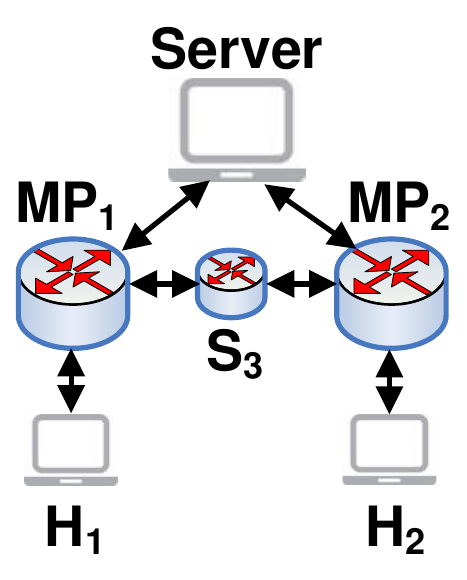}}
	\captionsetup{justification=centering}
  \caption{SW experiment setup.}
  \label{fig:P4Network}
  \end{subfigure}%

	\captionsetup{justification=raggedright}
  \caption{Experiment setup.}
  \label{fig:Setup}
\end{figure}

\ifdefined\LessText
\else
\begin{figure*}[htbp]
	\centering

  \begin{subfigure}[t]{.48\textwidth}
	\centering
  \fbox{\includegraphics[height=.5\textwidth]{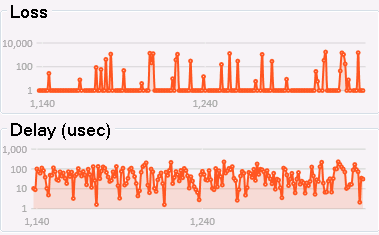}}
	\captionsetup{justification=centering}
  \caption{Occasional bursts.}
  \label{fig:TraceA}
  \end{subfigure}%
  \begin{subfigure}[t]{.48\textwidth}
	\centering
  \fbox{\includegraphics[height=.5\textwidth]{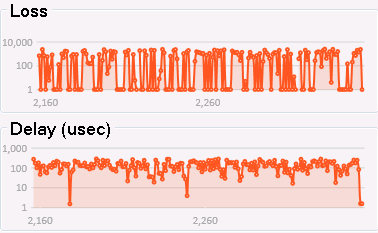}}
	\captionsetup{justification=centering}
  \caption{Frequent bursts.}
  \label{fig:TraceB}
  \end{subfigure}%

	\captionsetup{justification=raggedright}
  \caption{Experimental results from the hardware implementation. The horizontal axis represents the elapsed time since the beginning of the experiment in seconds. The vertical axis represents the number of packets lost per second (for `Loss'), and the delay between $MP_1$ and $MP_2$ in microseconds (for `Delay').}
  \label{fig:Traces}
\end{figure*}
\fi

\subsection{Hardware Implementation}
In this experiment we evaluated \ampm\ over a hardware switch. We used two development boards of a Marvell Prestera
switch device. The two devices were connected as illustrated in Fig.~\ref{fig:ExperimentNetwork}. The two switches were used as measurement points, and TimeFlip~\cite{Infocom-TimeFlip} was used in the switches in order to periodically toggle the color with an interval of 1~second. An off-the-shelf traffic generator was used for generating traffic through the two switches. The multiplexed marking method was used to measure loss and delay from $MP_1$ to $MP_2$, and the least significant bit of the DSCP field in the IPv4 header was used as the marking bit. The two MPs were synchronized using the Precision Time Protocol (PTP)~\cite{IEEE1588}, and periodically exported the counters and timestamps to a server that plotted the loss 
\ifdefined\LessText
and delay.
\else
and delay, as shown in Fig.~\ref{fig:Traces}. 
\fi

\ifdefined\LessText
\else
\textbf{Loss and delay under congestion.}
Traffic was run from the traffic generator through $MP_1$ and $MP_2$, and back to the traffic generator. Traffic was forwarded from $MP_1$ to $MP_2$ through a 10 Gbps link.
We started by generating traffic at a slightly higher rate than 10 Gbps, and observed the congestion between $MP_1$ and $MP_2$, indicated by the \ampm\ measurement.
The traffic flow from the traffic generator consisted of two sub-flows: (i) a 9.7 Gbps sub-flow with 64-byte packets, and (ii) a sub-flow with random-sized packets with an inter-packet gap of 21~microseconds. The size of the packets in the latter sub-flow was uniformly distributed between 64 and 1518 bytes. The random packet size caused congestion to be created in a non-deterministic manner, as illustrated in Fig.~\ref{fig:Traces}. Two scenarios are presented in the figure: occasional bursts are shown in (a), whereas frequent bursts were imposed in (b) by reducing the inter-packet gap to 20.95 microseconds. As the figures show, the loss rate and delay are correlated, and are higher when the monitored link is more congested (b). The delay, however is bounded by the queue size in $MP_1$, which was configured to ~300 microseconds in this experiment.
\fi

\begin{figure}[htbp]
	\centering

  \begin{subfigure}[t]{.24\textwidth}
	\centering
  \fbox{\includegraphics[height=.75\textwidth]{DemoDelay}}
	\captionsetup{justification=centering}
  \caption{Delay: measured vs. expected.}
  \label{fig:DemoDelay}
  \end{subfigure}%
  \begin{subfigure}[t]{.24\textwidth}
	\centering
  \fbox{\includegraphics[height=.75\textwidth]{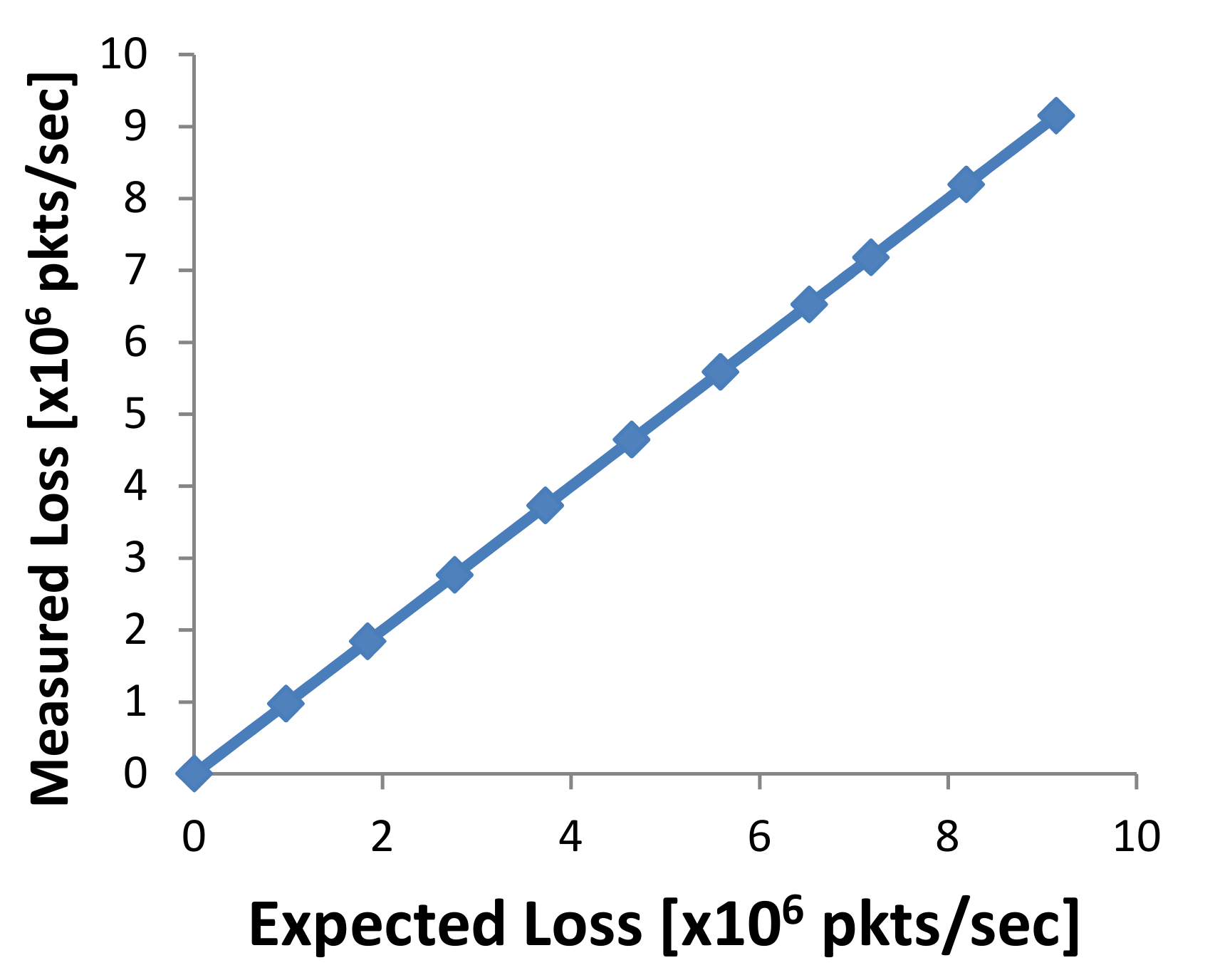}}
	\captionsetup{justification=centering}
  \caption{Packet loss: measured vs. expected.}
  \label{fig:DemoLoss}
  \end{subfigure}%

	\captionsetup{justification=raggedright}
  \caption{Experimental results of a hardware \ampm\ implementation.}
  \label{fig:DelayLoss}
\end{figure}

\ifdefined\LessText
In
\else
\textbf{Measurement accuracy.}
In the second part of 
\fi
this experiment we evaluated the accuracy of loss and delay measurement. In order to create synthetic packet loss in $MP_1$, we varied the traffic rate from the traffic generator to $MP_1$ from 10~Gbps to 15~Gbps,
thus creating a temporary overload of 0 to 50\%, respectively at the egress port of $MP_1$. We synthetically varied the delay by configuring the queue size in $MP_1$, and thus we were able to synthesize delay between 3~microseconds and 1~millisecond.

\ifdefined\LessText
\textbf{Results.}
\fi
The measurement results were monitored by a server that was attached to the two devices. The delay measurement error was less than 100~nanoseconds in all the tests, and the loss measurement error was zero in most measurements, and a single packet (less than .0001\% error) in some of the measurements. The loss and delay measurement results were as expected, confirming that \ampm\ can be implemented accurately over commodity switch silicons.

\subsection{P4 Software Implementation}
The purpose of this evaluation was to validate the simplicity of implementing \ampm\ in P4. Two variants of \ampm\ were implemented, the double marking method and the multiplexed marking method. The code is publicly available at~\cite{code}, and was implemented as two P4 applications running over the open source P4 switch from the P4 Consortium~\cite{P4Code}.

Our implementation includes an extension to the P4 reference switch that includes the time-of-day in the switch's intrinsic metadata. As discussed above, this extension allows multiple switches to be synchronized to common time intervals, enabling a muxed marking implementation along the lines of Section~\ref{MuxedSection}. It should be noted that it is possible to implement \ampm\ without clock synchronization,
\ifdefined\LessText
\footnote{As further discussed in the extended version of this paper~\cite{MarkingBasedTech}.} 
\else
\footnote{As further discussed in Section~\ref{SyncSec}.} 
\fi
and in this case the implementation can be purely in P4, without any changes to the existing P4 reference switch.

\begin{figure}[htbp]
	\centering

  \begin{subfigure}[t]{.24\textwidth}
	\centering
  \fbox{\includegraphics[width=.9\textwidth]{DelayResults}}
	\captionsetup{justification=centering}
  \caption{Delay: measured vs. configured.}
  \label{fig:DelayResults}
  \end{subfigure}%
  \begin{subfigure}[t]{.24\textwidth}
	\centering
  \fbox{\includegraphics[width=.9\textwidth]{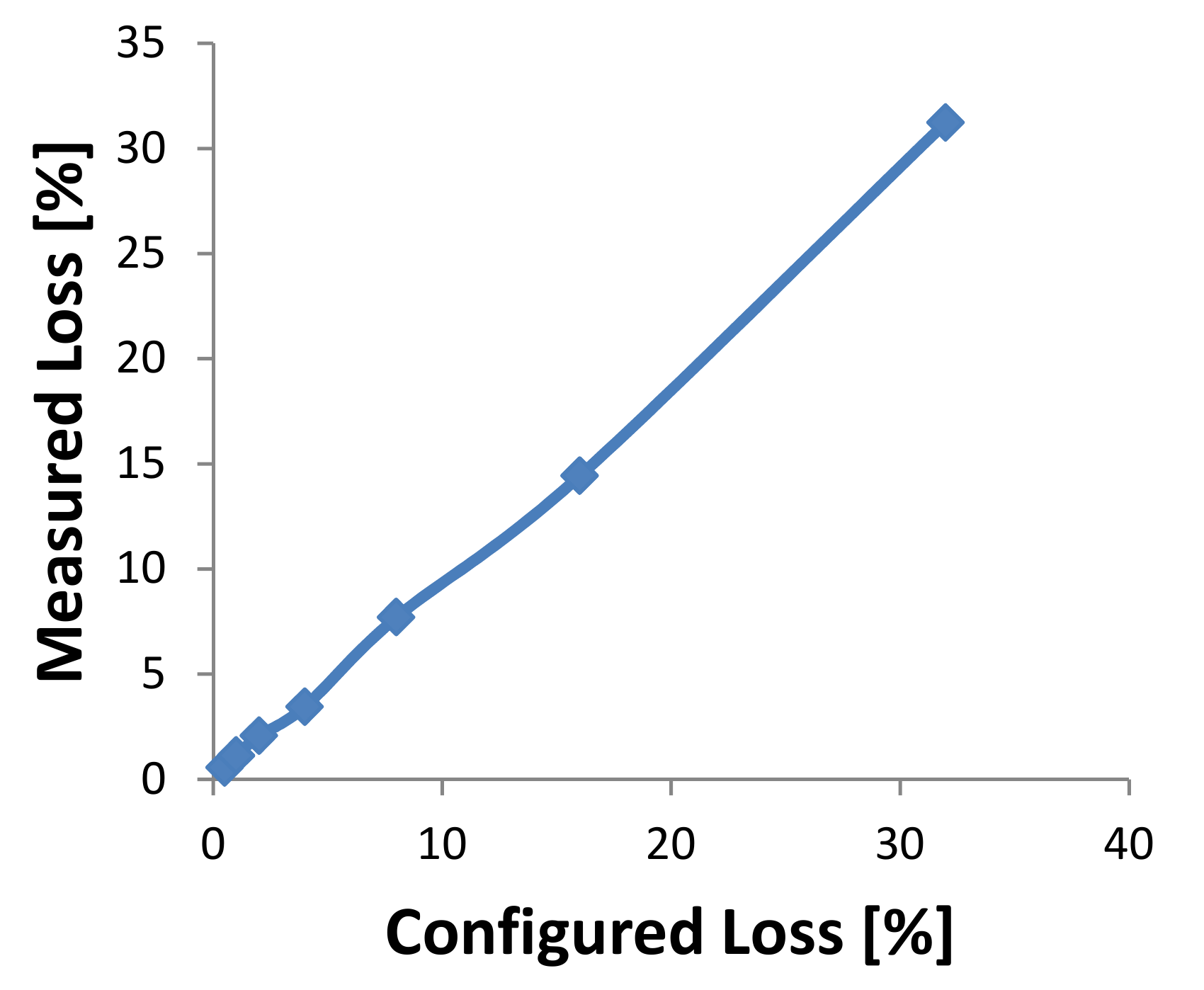}}
	\captionsetup{justification=centering}
  \caption{Packet loss: measured vs. configured.}
  \label{fig:LossResults}
  \end{subfigure}%

	\captionsetup{justification=raggedright}
  \caption{Experimental results of a P4-based \ampm\ implementation.}
  \label{fig:P4Results}
\end{figure}

The experiment was emulated in Mininet using the setup of Fig.~\ref{fig:P4Network}. Traffic was sent from host $H_1$ to host $H_2$ through the network, and was forwarded by switches $MP_1$, $S_3$ and $MP_2$. In the Mininet environment all switches use the Linux clock of the machine that hosts Mininet, and thus all switches are synchronized by definition. The two measurement points were $MP_1$ and $MP_2$, and \ampm\ was run with an interval of 16~seconds. 
\ifdefined\FivePage
\else
\footnote{Due to a limitation in our software implementation we did not test shorter measurement intervals.}
\fi
Results were collected by a third host, labeled `Server' in the figure. Loss and delay were synthetically configured in the Mininet environment in order to compare the measured performance and the expected performance.

\textbf{Results.}
The experimental results are summarized in Fig.~\ref{fig:P4Results}, and demonstrate the accuracy of the measurement compared to the anticipated loss and delay that were synthesized by Mininet. The delay measurement error was on the order of 1~millisecond compared to the anticipated delay, and the loss measurement error was on the order of 10\%. These results are reasonable given the software-based emulation environment that runs on a single machine.

\fi

\ifdefined\TwoPage
\else
\section{Related Work}
Performance measurement and monitoring protocols have been in deployment for many years; counters and statistics are often collected from network devices using SNMP~\cite{snmp} or NETCONF~\cite{netconf}, and various Operations, Administration and Maintenance (OAM) tools~\cite{mizrahi2014overview} are widely used in carrier networks. The literature is rich with publications about network telemetry and timestamping, e.g., ~\cite{narayana2017language,mittal2015timely,kimband}. \ampm\ was first introduced in~\cite{Tempia}, and has been under discussion in the IETF for a few years~\cite{AltMark}. A tutorial of \ampm\ ~\cite{AMPMCom} is currently under review. TimeFlip was first introduced in~\cite{Infocom-TimeFlip}, and~\cite{swfanDPT} showed that TimeFlip can be applied to Alternate Marking. The current paper is the first work that analyzes the programming abstractions and implementation considerations of marking-based telemetry, and presents experimental results from a P4-based implementation and from a silicon implementation. 

\section{Conclusion}
This paper analyzed two abstractions that enable the implementation of marking-based telemetry, and introduced a novel time-muxed parsing approach. Our implementation and experimental results demonstrate the simplicity of using these abstractions, and the accuracy and efficiency of marking-based telemetry. While time synchronization is not mandatory for implementing marking-based telemetry, we argue that implementations can benefit from using timestamps that are based on a globally synchronized time-of-day.

\section*{Acknowledgments}
We gratefully acknowledge Roy Mitrany for his dedicated help throughout this project. We also thank Yoram Revah for his great help in kicking off this work.

\fi

\ifdefined\FivePage
\pagebreak
\fi

\bibliographystyle{abbrv}
\bibliography{P4Alt}

\end{document}